\documentclass[12pt]{article}

\usepackage{enumitem}
\usepackage{float}
\usepackage{cite}
\usepackage{amsfonts}
\usepackage{amssymb}
\usepackage{amsmath}
\usepackage{xcolor}
\usepackage{mathtools}

\usepackage{graphicx}

\def\gtwid{\mathrel{\raise.3ex\hbox{$>$\kern-.75em\lower1ex\hbox{$\sim$}}}}
\def\ltwid{\mathrel{\raise.3ex\hbox{$<$\kern-.75em\lower1ex\hbox{$\sim$}}}}
\def\square{\kern1pt\vbox{\hrule height 1.2pt\hbox{\vrule width 1.2pt\hskip 3pt
   \vbox{\vskip 6pt}\hskip 3pt\vrule width 0.6pt}\hrule height 0.6pt}\kern1pt}

\begin{document}

\begin{titlepage}

\begin{flushright}
UFIFT-QG-24-07
\end{flushright}

\vskip 0cm

\begin{center}
{\bf The Other ADM}
\end{center}

\vskip 0cm

\begin{center}
R. P. Woodard$^{\dagger}$ and B. Yesilyurt$^{*}$
\end{center}

\vskip 0cm

\begin{center}
\it{Department of Physics, University of Florida,\\
Gainesville, FL 32611, UNITED STATES}
\end{center}

\vspace{0cm}

\begin{center}
ABSTRACT
\end{center}
In the peculiar manner by which physicists reckon descent, this article 
is by a ``child'' and ``grandchild'' of the late Stanley Deser. We begin 
by sharing reminiscences of Stanley from over 40 years. Then we turn to
a problem which was dear to his heart: the prospect that gravity might
nonperturbatively screen its own ultraviolet divergences and those of 
other theories. After reviewing the original 1960 work by ADM, we 
describe a cosmological analogue of the problem and then begin the 
process of implementing it in gravity plus QED.

\begin{flushleft}
\end{flushleft}

\vskip 0cm

\begin{flushleft}
$^{\dagger}$ e-mail: woodard@phys.ufl.edu \\
$^{*}$ e-mail: b.yesilyurt@ufl.edu
\end{flushleft}

\end{titlepage}

\section{Introduction}

Stanley Deser was the Grand Old Man of quantum gravity. Everyone 
in the field knew him, and the vast majority of us loved him. His 
life was a testament to the persistence of scientific inquiry, 
optimism and simple humanity over the course of a turbulent century.

Stanley was born in 1931 in interwar Poland. That wasn't a good
time or place to be Jewish. His immediate family just managed 
to escape the Nazis to the United States following the French 
collapse of 1940; most of his relatives did not. The sadly common
tragedy of those years left many scared, but it seems rather to 
have spurred Stanley to take full advantage his intellectual gifts.
He graduated college at 18, and took a Ph.D from Harvard at age 22.
In a career spanning seven decades he is credited with hundreds of
publications, including 8 papers and a book written after the
age of 90.

Stanley was trained in particle physics by Julian Schwinger but 
made the transition to gravity during a postdoc at the Institute 
for Advanced Study. Gravity needed him: despite the lovely 
geometrical formulation of its early days, general relativity was 
not then a proper field theory. There was no canonical formalism, 
with its careful enumeration of the degrees of freedom and their 
contribution to the total energy. Hence there was no way to prove 
classical stability \cite{Deser:1960zzb,Schoen:1979zz}, no way to 
develop numerical integration techniques \cite{Hahn:1964,Smarr:1977} 
and no way to even begin thinking about quantization 
\cite{Arnowitt:1959eec,DeWitt:1967yk}.

In a memorable sequence of papers \cite{Arnowitt:1959ah,Deser:1960zzc,
Arnowitt:1960es,Arnowitt:1961zz,Arnowitt:1962hi} with Dick Arnowitt 
and Charlie Misner, Stanley sorted out the gauge issue, identified 
canonical variables and defined an energy functional for asymptotically
flat geometries. Stanley would return to the problem of gravitational
energy later on in his career. With Claudio Teitelboim (now Bunster) 
he established the stability of supergravity \cite{Deser:1977hu}. And 
he collaborated with Larry Abbott to prove the classical stability of 
gravity with a positive cosmological constant \cite{Abbott:1981ff}.

Stanley was a consummate collaborator:
\begin{itemize}
\item{With David Boulware he showed that endowing the graviton with a
mass inevitably results in a ghost mode, provided that the theory has
a smooth perturbative limit \cite{Boulware:1972yco}.}
\item{With Peter van Nieuwenhuizen he extended the work of 't Hooft
and Veltman to show that renormalizability is lost at one loop when
general relativity is combined with either electromagnetism 
\cite{Deser:1974zzd,Deser:1974cz}, Yang-Mills theory \cite{Deser:1974nb,
Deser:1974xq}, or Dirac fermions \cite{Deser:1974cy}.}
\item{With Bruno Zumino he showed that consistently coupling a spin 
$\frac32$ gravitino to gravity produces a locally supersymmetric 
theory \cite{Deser:1976eh}. The two then applied their formalism to
string theory \cite{Deser:1976rb} and to the breaking of supersymmetry
\cite{Deser:1977uq}.}
\item{With Mike Duff and Chris Isham he identified the first true
conformal anomaly \cite{Deser:1976yx}, which led to a classification
scheme with Adam Schwimmer \cite{Deser:1993yx}.}
\item{With Roman Jackiw and Stephen Templeton he showed that adding a 
dimensionally-reduced Chern-Simons term to Yang-Mills or gravity 
results in massive particles of spin 1 and 2, respectively 
\cite{Deser:1981wh,Deser:1982vy}.}
\item{With Gerard 't Hooft and Roman Jackiw he explained how to 
understand general relativity in $2 + 1$ dimensions \cite{Deser:1983tn}.}
\item{With Cedric Deffayet and Gilles Esposito-Farese he extended flat
space Galileons to curved space \cite{Deffayet:2009mn}.}
\end{itemize}

Stanley Deser was also an inspiring teacher, mentor and friend to young
researchers in quantum gravity. This article is by one of his students 
(RPW) and by that person's own student (BY). After this brief introduction, 
Stanley's student shares some personal recollections from 40 years of 
association. Then the two authors turn to a problem which long fascinated 
Stanley: the possibility that quantum gravity might regulate its own 
ultraviolet divergence problem and even those of other theories 
\cite{Deser:1957zz}. We first review ADM's prescient and thought-provoking
demonstration that classical gravitation cancels the self-energy 
divergence of a point charge \cite{Arnowitt:1960zza,Arnowitt:1960zz}.
This is ``the other ADM'' of our title. We then describe a quantum field
theoretic analogue of the same basic effect in the context of inflationary
cosmology. The next section discusses the prospects for implementing the 
ADM mechanism in general relativity plus QED on an asymptotically flat
background. Our conclusions comprise the final section.

\section{Remembering Stanley}

I was a graduate student at Harvard from the fall of 1977 through the
spring of 1983. Harvard theorists of that era were quite hostile to
quantum gravity, as strange as that seems in view of the current 
direction of research at that place. Whenever a Harvard graduate 
student aspired to work on the subject he was required to do it
through a sort of underground railroad in which Sidney Coleman served
as a front for Stanley Deser. Lee Smolin took that route before me, and 
Paul Renteln \cite{Renteln:1988wi} came afterwards.

In 1981 my lifelong friend, Nick Tsamis, and I conceived the notion 
of generalizing gravitational Green's functions to invariants based on
the same procedure Stanley Mandelstam had employed for gauge theories
\cite{Mandelstam:1968hz}. Our Green's functions involved products of 
Riemann tensors, evaluated at the ends of operator-valued geodesics from a 
common origin, with their tensor indices parallel-transported back to the 
origin and contracted into vierbeins \cite{Woodard:1983bh}. With a measure 
factor at the origin one can prove that these things are invariants when 
evaluated in a Poincar\'e invariant gauge \cite{Tsamis:1984kx,
Tsamis:1989yu}. Stanley chanced to be at CERN that year so I pitched the 
idea in a long letter. The reply came back in what I would later recognize 
as his typically laconic style: ``You understand field theory.'' Stanley 
assured Sidney that my project with Nick would make a reasonable thesis, 
and we set out to compute the invariant 2-point function at 1-loop order.

The calculation consumed over a year. Every few weeks I would take the 
commuter train from Cambridge to Waltham to show Stanley our progress. 
At that time he smoked a pipe and resided in a cozy corner office of 
Physics building on the Brandeis campus. There was a cartoon outside his 
door, depicting a distinguished-looking figure, not unlike Stanley, and 
bearing the caption ``Classical Physicist''. Stanley was a speed-reader 
who would flip quickly pages of intricate calculations, and later of my 
draft thesis. I didn't see how anyone could absorb information at that 
rate but he would periodically stop and draw attention to something which 
required attention.

A physicist becomes very vulnerable when he undertakes a long computation
which will show few results until the end. Stanley saw to it that I had 
the necessary support and protection. When it came time to apply for 
postdoc positions and I was still mired in the last stages of debugging, 
he took up the slack. I might have taken off a week to type up a dozen 
applications; Stanley did the rest. I could not have asked for a better 
first postdoc than the one he secured for me with Bryce DeWitt at the UT 
Austin. During the summer of 1983, with my over 600-page thesis still 
incomplete, Stanley used his own grant money to support me at Brandeis 
so I wouldn't need to teach. I was often amazed at the extent of his 
support --- he even offered to come to Austin to hear my final defense 
--- but I was never disappointed.

Although Stanley and I co-authored a paper in 2019 \cite{Deser:2019lmm},
and corresponded almost to the end of his life, the last time I actually 
saw him was in May of 2015. My wife and I were attending a program at
KITP in Santa Barbara. We both had to fly back from LAX to Taiwan, but 
I was to leave a week before her in order to substitute-teach for her
graduate EM course. Deser invited us to give a talk at Caltech. I baked 
a German chocolate cake, rented a car and we drove to Pasadena, planning 
to afterwards drop the car at LAX and catch my late night flight while 
my wife took the last Santa Barbara airbus back. Kelly Stelle joined us 
for a memorable dinner at the Atheneum. We enjoyed ourselves so much 
that I almost missed my flight, and my wife did miss her airbus. With
characteristic thoughtfulness, Stanley arranged to cover her hotel bill.

Despite of being a power in the world of gravity, Stanley kept a 
remarkably low profile. After decades of acquaintance I had not known 
he chaired the NSF committee which initially recommended funding LIGO. 
Nor did I learn this from Stanley; Rainer Weiss recounted the story 
at the ADM-50 conference in 2009. Another well-deserved intervention
was the role Stanley played in getting Thibault Damour his position
at IHES; my friend Nick passed this along after Thibault told him. 
These tales are legion. Kelly Stelle commented that Stanley had
done so much for so many people that there would be a crowd at the 
2004 festschrift we organized \cite{Liu:2006sx}. Characteristically, 
it was only through the intervention of Stanley's wife Elsbeth that 
he acquiesced to the event in his honor.

I have known some of the greatest minds of theoretical physics
in the past few decades. Many of them are not very nice people, and 
I struggle to overlook the boorishness and cruelty which too often 
demean scientific genius. I never had to struggle with Stanley; he 
enriched the world's soul along with its store of knowledge. He was 
like a father to me, and I miss him as one mourns the loss of a parent.

Stanley got along even with people who were famously difficult. I never 
heard him utter an unkind word; the most he ever did was to gently poke 
fun at certain people. For example, he referred to a physicist who had 
claimed to solve the problem of quantum gravity as ``the pride of
XXXX.'' And we kidded one another about the address another off-mass 
shell colleague would give at ADM-100 after winning the Nobel Prize.

Stanley was as loyal and solicitous to his own advisor,
Julian Schwinger, as I have tried to be to him. Schwinger was
famously shy, which led to some amusing incidents. In 1985 Stanley 
arranged for me to have dinner with Schwinger and his wife in 
their Bel-Air home as part of a postdoc interview. The candidate 
for such a position would normally have been invited to give a talk 
at UCLA, but Schwinger felt more secure at home, and I was of course 
honored to meet the man who had done so much for quantum field theory. 
Later, Stanley invited Schwinger to visit the ITP while he was in 
charge of a workshop there. Schwinger declined on the grounds that 
people would expect him to give a talk, whereupon Stanley assured 
him that no talk was necessary if that was a problem. Schwinger 
demurred again, this time because people would expect him to show up 
at the office. Stanley, who was by then accustomed to these sorts of
objections, implored his mentor to visit and not show up at the office!

Although I could never be as supportive of my own students as
Stanley was to me, I like to believe that we become close. Still, 
I was astonished and humbled at being asked to name the firstborn
child of Changlong Wang (who worked on a project suggested by
Stanley \cite{Wang:2014tza,Wang:2015eaa}) and his wife, Qianqian 
Huang. I never had any doubt whose name the child should receive. 
Nor do I doubt that he has a great future based on an incident from 
Changlong's graduation. The UF's custom is that doctoral recipients 
are escorted by their advisors, decked out in the academic regalia 
of their own graduate schools. I asked Qianqian how she liked the 
Orange and Blue of her husband's robes and she demurely replied 
that she preferred my crimson and planned for her children to wear 
it one day. Both parents are brilliant, and fanatically hard 
workers, so I wouldn't bet against it!
\begin{figure}[H]
\centering{\includegraphics[height=5cm,width=6cm]{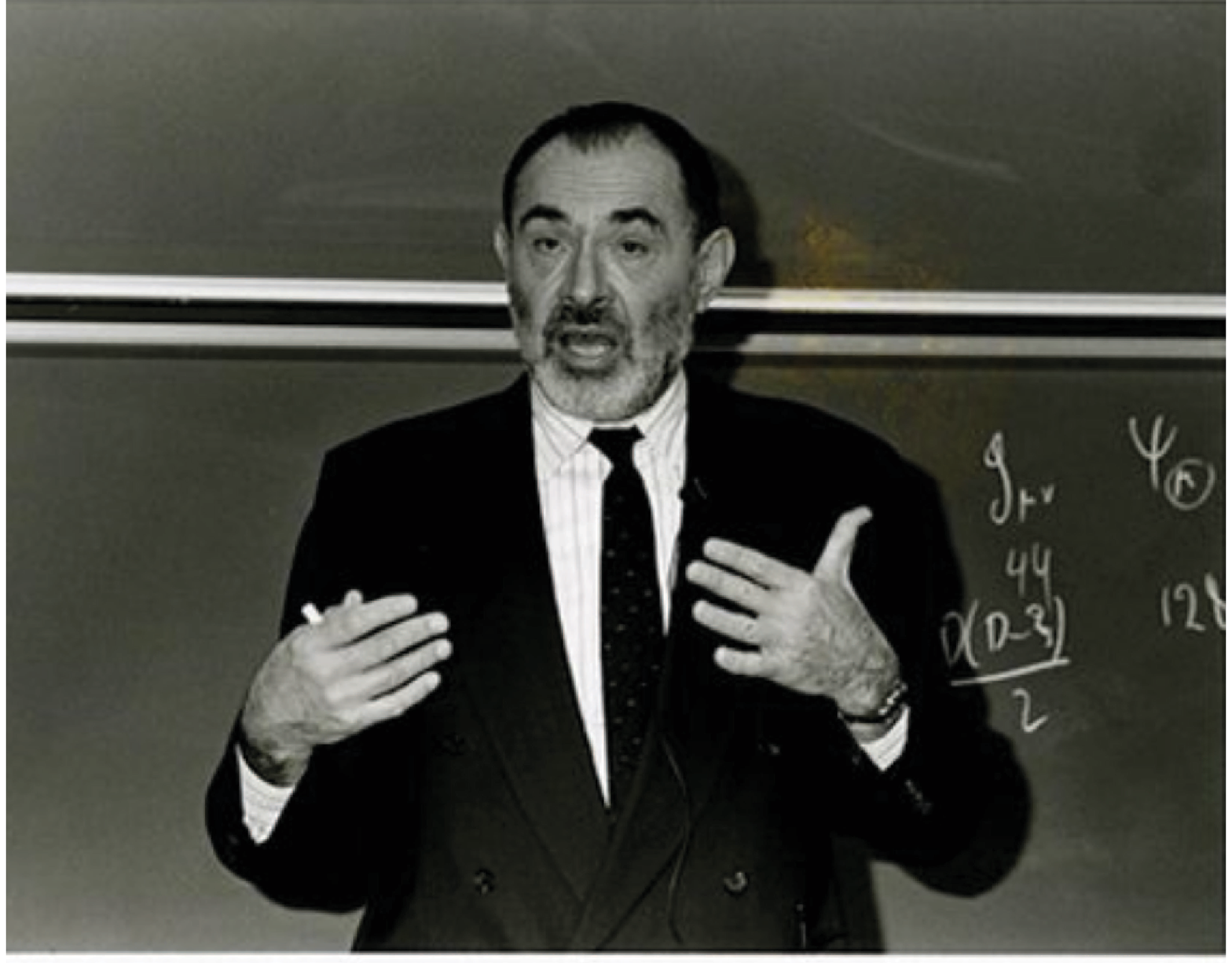}
\includegraphics[height=5cm,width=6cm]{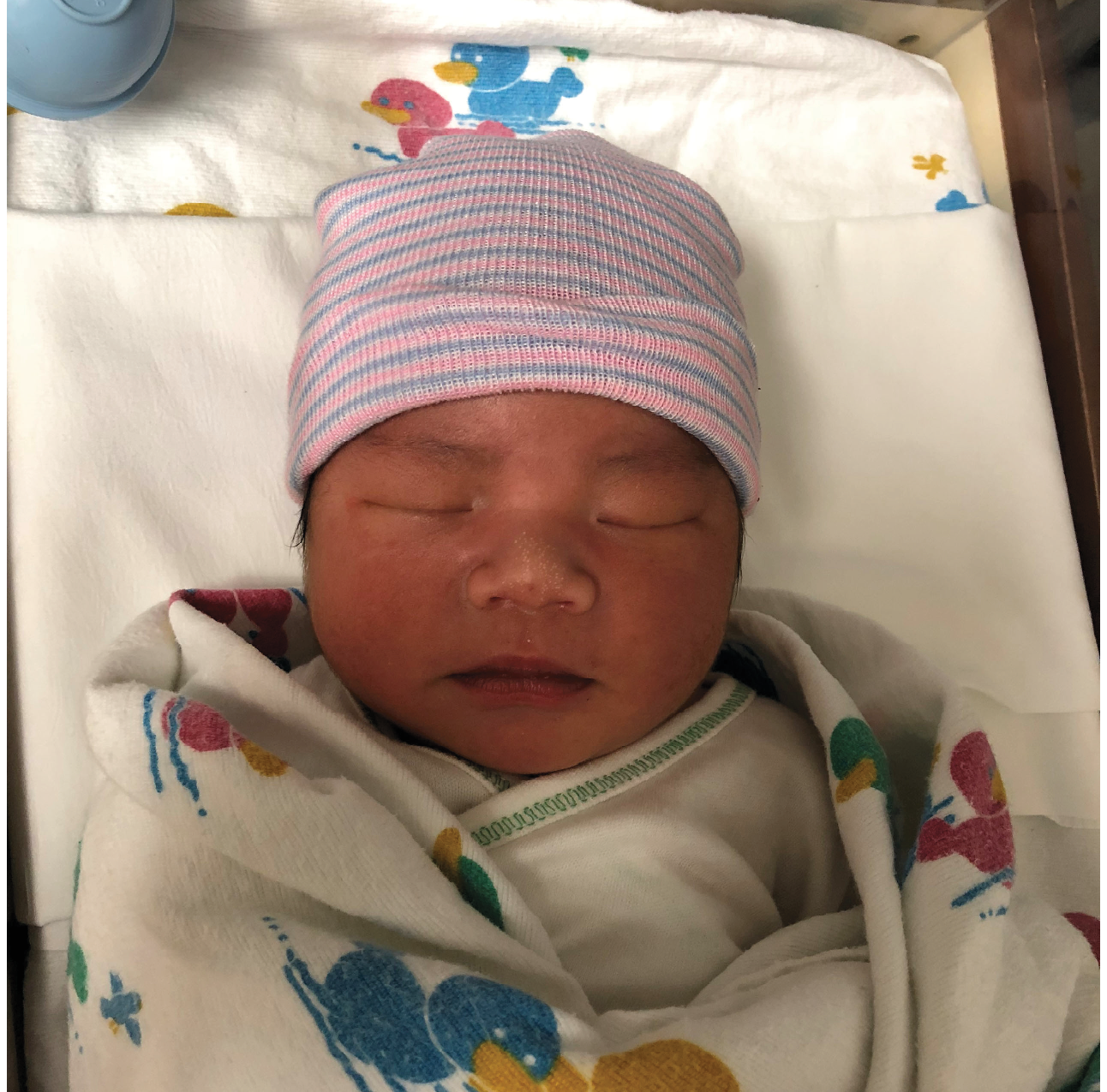}}
\caption{Stanley Deser and namesake, Stanley Wang, born May 
24, 2018.}
\label{BabyStan}
\end{figure}

In addition to Schwinger, Stanley had a special relationship with
Steven Weinberg. He was the one who recommended that Weinberg consult
me about the Schwinger-Keldysh formalism \cite{Weinberg:2005vy,
Weinberg:2006ac}, which was hugely important for my career. The only 
time I recall Stanley being angry was when a blogger criticized 
Weinberg. And Stanley was the driving force behind our nomination
of Weinberg for the Breakthrough Award. I don't know if our efforts 
had anything to do with 2020 award; the committee did not use the 
citation we proposed, but perhaps we planted the seed.

Stanley had a profound faith that truth will win out in the end.
However, that doesn't mean he always accepted the consensus views 
of physics at any time. He was particularly suspicious about the 
existence of dark matter and dark energy. I share his skepticism.
It strains credulity to believe that more than 95\% of the current 
energy density consists of exotic matter which we have never seen 
in the laboratory. As long as the dark sector can only be detected 
through its gravity, modified gravity would seem to be an equally 
plausible explanation. Stanley and I proposed that this might be 
accomplished by nonlocal modifications of gravity, derived from 
secular effects during a prolonged epoch of inflationary particle
production, which grow nonperturbatively strong \cite{Deser:2007jk,
Deser:2013uya,Deser:2019lmm}. Our model was designed to explain 
the current epoch of cosmic expansion but Stanley's collaborators,
Cedric Deffayet and Gilles Esposito-Farese, worked with me to devise
a model which explains gravitationally bound structures without 
dark matter \cite{Deffayet:2011sk,Deffayet:2014lba}. It used to
be said that modified gravity models cannot account for cosmological
perturbations but Cedric and I have just devised a model which
does that \cite{Deffayet:2024ciu}. What is more, Shun-Pei Miao, 
Nick Tsamis and I are getting ever closer to {\it deriving} such 
models from first principles \cite{Miao:2024nsz,Miao:2024shs}. If 
we succeed, it will vindicate another of Stanley's visions.

Although English was not his native language, Stanley wrote and
spoke it with great erudition. He once joked about having lost 
track of how many tongues he had learned and then forgotten, 
however, the French of his childhood stayed to the end. I 
happened to be visiting IHES during his final illness, in April 
of 2023. Upon hearing the news about his old friend's condition, 
Thibault Damour e-mailed Stanley and was relieved (prematurely, 
as it turned out) to receive a reply in perfect French.

I have already mentioned Stanley's penchant for brevity. He maintained 
that too much explanation in a paper or a talk, or even in correspondence,
insulted the intellect of the audience. After I recounted the perilous 
state of Egypt, a few years before the Arab Spring, Stanley's reply was, 
``Clever of the Jews to get out.'' In my naivete I once cited a rival's 
review paper, to which Stanley confined himself to observing, ``That's not 
my favorite reference.'' Whenever I bemoaned the inexplicable fads which 
dominate particle physics, he would invariably reply, ``Physics is what 
physicists do.'' The longest comment I recall came as he was paring down 
the prose in my first draft of a joint paper. He jokingly accused me of 
``including everything but the kitchen sink --- whereupon you threw that 
in too!'' My last note from him formed a fitting coda to 40 years of 
correspondence. I had wished him a happy 92nd birthday and mentioned an 
up-coming trip to France to work with our mutual friend Cedric Deffayet. 
His reply was inimitable: ``thanks!still in bed but alive.happy travels 
sd''. Rest in peace Stanley.

\section{The ADM Mechanism}

The idea that gravity might regulate divergences is based on the fact
that gravitational interaction energy is negative \cite{Deser:1957zz,
DeWitt:1964yh}. For example, this makes the mass of the Earth-Moon system 
slightly smaller than the sum of their masses, even when one includes 
the kinetic energy of their orbital motion,
\begin{equation}
M_{\rm EM} = M_{\rm E} + M_{\rm M} - \frac{G M_{\rm E} M_{\rm M}}{2 c^2 R}
\; . \label{EarthMoon}
\end{equation}
The decrease works out to about $6 \times 10^{11}$ kilograms, which
makes for a fractional reduction of $10^{-13}$. Note that the fractional
reduction becomes larger as the orbital radius $R$ decreases. 

In 1960 Arnowitt, Deser and Misner quantified the mechanism in the
context of a classical (in the sense of non-quantum) charged and
gravitating point particle \cite{Arnowitt:1960zza,
Arnowitt:1960zz}.\footnote{Their article in Physical Review Letters 
\cite{Arnowitt:1960zz} incidentally marks the first appearance of general
relativity in that journal, which is a measure of how much things have 
changed since then.} Although they solved the full general relativistic 
constraints and then computed the ADM mass, their result can be 
understood using a simple model that they devised. Suppose the particle
has a bare mass $M_0$ and charge $Q$, and is regulated as a spherical 
shell of radius $R$. Then its rest mass energy might be expressed as,
\begin{equation}
M(R) c^2 = M_0 c^2 \!+\! \frac{Q^2}{8\pi \epsilon_0 R} \!-\! 
\frac{G M^2(R)}{2 R} \; , \label{ADM1}
\end{equation}
where the single concession to relativity is that the Newtonian 
gravitational interaction energy has been evaluated using the total
mass. Of course the quadratic equation (\ref{ADM1}) can be solved
to give,
\begin{equation}
M(R) = \frac{c^2 R}{G} \Biggl[ \sqrt{1 \!+\! \frac{2 G M_0}{R c^2} 
\!+\! \frac{G Q^2}{4 \pi \epsilon_0 R^2 c^4} } - 1 \Biggr] \; . 
\label{ADM2} 
\end{equation}
The unregulated limit is finite and independent of the bare mass,
\begin{equation}
\lim_{R \rightarrow 0} M(R) = \sqrt{ \frac{Q^2}{4\pi \epsilon_0 G}} 
\; . \label{ADM3}
\end{equation}

Three crucial points about the result (\ref{ADM3}) deserve mention:
\begin{itemize}
\item{It is finite;}
\item{It is independent of the bare mass $M_0$, as long as that is 
finite; and}
\item{It is nonperturbative.}
\end{itemize}
Of course finiteness results from the fact that gravitational 
interaction energy is negative. This is evident from expression 
(\ref{ADM1}). The $Q^2/8\pi\epsilon_0 R$ term means that compressing 
a shell of charge costs energy, however, the $-G M^2/2R$ term signals
that gravity is able to pay the bill, no matter how high.

The fact that any fixed $M_0$ drops out is also evident from expression 
(\ref{ADM1}). Note that this is not at all how a conventional particle 
physicist would have approached the problem. Our conventional colleague 
would have regarded the total mass $M$ as a measured quantity and then 
required the bare mass to depend upon the regulating parameter $R$ so 
as to force the result to agree with measurement,
\begin{equation}
M_0(R) = M_{\rm meas} - \frac{Q^2}{8\pi \epsilon_0 R c^2} + \frac{G
M_{\rm meas}^2}{2 R c^2} \; . \label{renorm}
\end{equation}
That is how renormalization works. It is unavoidable without gravity,
but the presence of gravity opens up the fascinating prospect of 
{\it computing fundamental particle masses from first principles}. 
Setting $Q = e$ in expression (\ref{ADM3}) gives an impossibly large
result for the electron,
\begin{equation}
\sqrt{\frac{e^2}{4 \pi \epsilon_0 G}} = \sqrt{\frac{e^2}{4 \pi \epsilon_0 
\hbar c}} \times \sqrt{\frac{\hbar c}{G}} = \sqrt{\alpha} \times 
M_{\rm Planck} \; . \label{classical}
\end{equation}
However, it is well known that quantum field theoretic effects soften
the linear self-energy divergence of a classical electron to a logarithmic 
divergence \cite{Weisskopf:1939zz}. This open the possibility of the true
relation containing exponentials. For example, one gets within a factor 
of four with,
\begin{equation}
M_{\rm electron} = \sqrt{\alpha} M_{\rm Planck} \times \exp\Bigl[-
\frac1{e^1 \alpha}\Bigr] \approx 0.134~{\rm MeV} \; ,
\end{equation}
where $e^1 \approx 2.71828$ is the base of the natural logarithm. The 
electron also carries weak charge, which should enter at some level. 
Perhaps all fundamental particle masses can be computed from first 
principles? One might even hope that the mysterious generations of the 
Standard Model appear as ``excited states'' in such a picture.

The nonperturbative nature of the ADM mechanism is evident from the
fact that (\ref{ADM3}) goes like the square root of the fine structure
constant and actually diverges as Newton's constant goes to zero. The
perturbative result comes from expanding the square root of (\ref{ADM2})
in powers of $Q^2$ and $G$,
\begin{eqnarray}
\lefteqn{M(R) = \Bigl( M_0 \!+\! \frac{Q^2}{8\pi \epsilon_0 R c^2}\Bigr) 
\Biggl\{1 - \frac14 \Bigl(\frac{2 G M_0}{R c^2} \!+\! \frac{G Q^2}{4\pi 
\epsilon_0 R^2 c^4}\Bigr) } \nonumber \\
& & \hspace{1cm} + \frac18 \Bigl(\frac{2 G M_0}{R c^2} \!+\! 
\frac{G Q^2}{4\pi \epsilon_0 R^2 c^4} \Bigr)^2 -\frac{5}{64} 
\Bigl(\frac{2 G M_0}{R c^2} \!+\! \frac{G Q^2}{4\pi \epsilon_0 R^2 c^4} 
\Bigr)^3 + \ldots\Biggr\} . \qquad \label{ADM4}
\end{eqnarray}
This is a series of ever-higher divergences. Of course perturbation 
theory becomes invalid for large values of the expansion parameter,
\begin{equation}
\frac{2 G M_0}{R c^2} + \frac{G Q^2}{4\pi\epsilon_0 R^2 c^4} \; . 
\label{parameter}
\end{equation}
Perhaps the same problem invalidates the use of perturbation theory in
quantum general relativity, which would show cancellations like 
(\ref{ADM3}) if only we could devise a better approximation scheme? 

It is hopeless trying to perform a genuinely nonperturbative computation.
However, a glance at the expansion (\ref{ADM4}) shows what goes wrong 
with conventional perturbation theory: gravity has no chance to ``keep 
up'' with the gauge sector. The lowest electromagnetic divergence is
$Q^2/8\pi\epsilon_0 R c^2$, whereas gravity's first move to cancel 
comes at order $-[Q^2/4\pi\epsilon_0 R c^2]^2 \times G/8R c^2$. What is
needed is a reorganization of perturbation theory in which the 
gravitational response comes at the ``same order'' as the gravitational
response. Several studies have searched for such an expansion without 
success \cite{Duff:1973zz,Isham:1970aw,Woodard:1997dm,Casadio:2009eh,
Casadio:2009jc,Mora:2011jj}.

\section{A Cosmological Analogue}

Two additional points should be noted concerning the ADM result 
(\ref{ADM3}):
\begin{itemize}
\item{The regularization of divergences arises from solving 
``Hamiltonian constraint'' (\ref{ADM1}); and}
\item{The mass vanishes for $Q=0$, which means that gravity will
cancel {\it everything} in the absence of some sort of nongravitational
``charge''.}
\end{itemize}
The two limitations of the ADM analysis are taking the classical
limit and suppressing almost all the degrees of freedom. The theory
of primordial perturbations \cite{Starobinsky:1979ty,Mukhanov:1981xt,
Mukhanov:1990me} furnishes an example in which the Hamiltonian 
constraint can be solved without either taking the classical limit or 
suppressing any physical degree of freedom.

The background geometry of primordial inflation is characterized by a 
scale factor $a(t)$, Hubble parameter $H(t)$ and first slow roll
parameter $\epsilon(t)$,
\begin{equation}
d\overline{s}^2 = -dt^2 + a^2(t) d\vec{x} \!\cdot\! d\vec{x} \qquad ,
\qquad H(t) \equiv \frac{\dot{a}}{a} \quad , \quad \epsilon(t) \equiv 
-\frac{\dot{H}}{H^2} \; . \label{geometry}
\end{equation}
Inflation is characterized by positive first and second derivatives of 
the scale factor, that is, $H > 0$ and $0 \leq \epsilon < 1$. The 
simplest model consists of general relativity plus a minimally coupled 
scalar inflaton $\varphi$ whose slow roll down its potential 
$V(\varphi)$ provides the stress-energy of inflation,
\begin{equation}
\mathcal{L} = \frac{R \sqrt{-g}}{16 \pi G} - \frac12 \partial_{\mu} 
\varphi \partial_{\nu} \varphi g^{\mu\nu} \sqrt{-g} - V(\varphi) 
\sqrt{-g} \; . \label{infL}
\end{equation}
The scalar background and its potential are related to the geometrical
parameters (\ref{geometry}),
\begin{equation}
\dot{\varphi}_0^2 = -\frac{\dot{H}}{4\pi G} \qquad , \qquad V(\varphi_0)
= \frac{(\dot{H} \!+\! 3 H^2)}{8\pi G} \; . \label{infback}
\end{equation}

It is usual to describe the full metric using the ADM parameterization 
\cite{Arnowitt:1959ah},
\begin{equation}
ds^2 = -N^2 dt^2 + \gamma_{ij} \Bigl(dx^i \!-\! N^i dt\Bigr) \Bigl(dx^j 
\!-\! N^j dt\Bigr) \; . \label{ADMmetric}
\end{equation}
The $3+1$ decompositions of the metric and its inverse are,
\begin{equation}
g_{\mu\nu} = \left( \begin{matrix} 
-N^2 + \gamma_{k\ell} N^k N^{\ell} & -\gamma_{n \ell} N^{\ell} \\
-\gamma_{mk} N^k & \gamma_{k\ell}
\end{matrix} \right) \quad , \quad 
g^{\mu\nu} = \left( \begin{matrix}
-\frac1{N^2} & -\frac{N^n}{N^2} \\
-\frac{N^m}{N^2} & \gamma^{mn} - \frac{N^m N^n}{N^2} 
\end{matrix} \right) \; . \label{3+1}
\end{equation}
ADM also showed that a Lagrangian of the form (\ref{infL}) depends
upon the lapse $N(t,\vec{x})$ in a very simple way which is familiar
from elementary mechanics \cite{Arnowitt:1959ah},
\begin{equation}
\mathcal{L} = \Bigl({\rm Surface\ Terms}\Bigr) +
\frac{\sqrt{\gamma}}{16\pi G} \Bigl( \frac{K}{N} - N \!\cdot\! P 
\Bigr) \; . \label{Nform}
\end{equation}
The quantity $K$ can be recognized as a kinetic energy,
\begin{equation}
K = E_{ij} E_{k\ell} \Bigl( \gamma^{ik} \gamma^{j\ell} - \gamma^{ij}
\gamma^{k\ell}\Bigr) + 8\pi G \Bigl(\dot{\varphi} - \varphi_{,i} N^i
\Bigr)^2 \; , \label{kinetic}
\end{equation}
where $E_{ij}/2N$ is the extrinsic curvature,
\begin{equation}
E_{ij} \equiv \frac12 \Bigl( -\dot{\gamma}_{ij} + N_{i ; j} +
N_{j ; i}\Bigr) \; . \label{extrinsic}
\end{equation}
Here $N_{i} \equiv \gamma_{ik} N^{k}$ and $N_{i ; j}$ represents its
covariant derivative with respect to the 3-metric $\gamma_{ij}$. Note
that the gravitational kinetic energy from the trace part of 
$\gamma_{ij}$ is negative. The quantity $P$ consists of spatial 
gradient and potential energy,
\begin{equation}
P = -\mathcal{R} + 16 \pi G \Bigl( \frac12 \gamma^{ij} \partial_i 
\varphi \partial_j \varphi + V(\varphi) \Bigr) \; . \label{potential}
\end{equation}
where $\mathcal{R}$ is the Ricci scalar formed from the 3-metric
$\gamma_{ij}$.

The 3-metric is written in terms of a scalar perturbation 
$\zeta(t,\vec{x})$ and a traceless graviton field $h_{ij}(t,\vec{x})$,
\begin{equation}
\gamma_{ij}(t,\vec{x}) \equiv a^2(t) \!\times\! e^{2 \zeta(t,\vec{x})} 
\!\times\! \Bigl[e^{h(t,\vec{x})}\Bigr]_{ij} \qquad , \qquad 
h_{ii}(t,\vec{x}) = 0 \; . \label{infmetric}
\end{equation}
Although ADM fixed the gauge by choosing $N(t,\vec{x})$ and $N^i(t,\vec{x})$,
most cosmologists instead impose the conditions \cite{Maldacena:2002vr,
Weinberg:2005vy},
\begin{equation}
\varphi(t,\vec{x}) = \varphi_0(t) \qquad , \qquad \partial_j 
h_{ij}(t,\vec{x}) = 0 \; . \label{gauge}
\end{equation}
This converts the lapse and shift into functionals of the dynamical 
variables $\zeta$ and $h_{ij}$ which are determined by the Hamiltonian
and Momentum constraints,
\begin{eqnarray}
0 = \frac{\delta S}{\delta N} &\!\!\! = \!\!\!& -\frac{\sqrt{\gamma}}{16 \pi G}
\Bigl[ \frac{K}{N^2} + P\Bigr] \; , \label{Hamiltonian} \\
0 = \frac{\delta S}{\delta N^i} &\!\!\! = \!\!\!& \frac{\sqrt{\gamma}}{8 \pi G}
\Bigl[ \partial_i \Bigl(\frac{E}{N}\Bigr) - \Bigl( \frac{E_{ij}}{N}\Bigr)^{;j}
\Bigr] \; . \qquad \label{Momentum}
\end{eqnarray}
The Momentum Constraint (\ref{Momentum}) can only be solved perturbatively
in the weak fields $\zeta$ and $h_{ij}$,
\begin{equation}
N^i = \frac{\partial_i}{\nabla^2} \Bigl( \epsilon \zeta - \frac{\nabla^2}{a^2}
\, \frac{\dot{\zeta}}{H} \Bigr) + \dots \label{shift}
\end{equation}
However, the Hamiltonian Constraint (\ref{Hamiltonian}) can be solved exactly, 
both for the lapse and for the gauge-fixed, constrained Lagrangian 
\cite{Kahya:2010xh},
\begin{equation}
N = \sqrt{-\frac{K}{P}} \qquad \Longrightarrow \qquad \mathcal{L} 
\longrightarrow -\frac{\sqrt{\gamma}}{8 \pi G} \sqrt{-K P} \; .
\label{constrained}
\end{equation}

The nonperturbative solution of the Hamiltonian Constraint (\ref{constrained})
is what we have been seeking. No one knows its impact on the ultraviolet 
problem but the weak field expansion of the gauge-fixed and constrained 
Lagrangian does show an ADM-like erasure of scalar perturbation $\zeta$.
The quadratic terms are,
\begin{equation}
\mathcal{L} \longrightarrow \frac{a^3 \epsilon}{8\pi G} \Bigl[
\dot{\zeta}^2 - \frac{\partial_k \zeta \partial_k \zeta}{a^2} \Bigr] 
+ \frac{a^3}{64 \pi G} \Bigl[ \dot{h}_{ij} \dot{h}_{ij} -
\frac{\partial_k h_{ij} \partial_k h_{ij}}{a^2} \Bigr] + \Bigl({\rm 
Interactions}\Bigr) \; . \label{constL}
\end{equation}
Note from (\ref{infL}) that the scalar perturbation had unit strength 
before imposing the constraints, even after gauge fixing (\ref{gauge}). 
The gravitational constraints have almost completely erased it at the 
quadratic level (\ref{constL}); it is protected only by the (very small)
global ``charge'' associated with a nonzero first slow roll parameter 
$\epsilon$ of the background (\ref{geometry}). This suppression is not
an artifact of the quadratic action. Analysis of higher constrained 
interactions reveals that an extra factor of $\epsilon$ arises for each 
one or two extra powers of $\zeta$\cite{Maldacena:2002vr,Seery:2006vu,
Jarnhus:2007ia,Xue:2012wi}.

\section{Implementation in QED $+$ GR}

In this section we begin implementing the solution of the Hamiltonian
Constraint for QED $+$ general relativity. The dynamical variables are
the spacelike metric $g_{\mu\nu}$, the electromagnetic vector potential
$A_{\mu}$ (with field strength $F_{\mu\nu} \equiv \partial_{\mu} A_{\nu}
- \partial_{\nu} A_{\mu}$) and the Dirac bispinor $\psi_i$. The 
Lagrangian is,
\begin{eqnarray}
\lefteqn{ \mathcal{L} = \frac{R \sqrt{-g}}{16 \pi G} - \frac14 
F_{\rho\sigma} F_{\mu\nu} g^{\rho\mu} g^{\sigma\nu} \sqrt{-g} }
\nonumber \\
& & \hspace{2.3cm} + \overline{\psi} e^{\mu}_{~a} \gamma^{a} \Bigl(
i\partial_{\mu} - e A_{\mu} - \frac12 A_{\mu bc} J^{bc}
\Bigr) \psi \sqrt{-g} - m_0 \overline{\psi} \psi \sqrt{-g} \; . 
\qquad \label{GR+QED}
\end{eqnarray}
The vierbein field $e^{\mu}_{~a}$ is not an independent variable but 
rather a function of the metric determined by a local Lorentz gauge 
condition (about which more later) and the relation,
\begin{equation}
g^{\mu\nu} = e^{\mu}_{~a} e^{\nu}_{~b} \eta^{ab} \; . \label{vierbein}
\end{equation}
The spin connection is formed from it,
\begin{equation}
A_{\mu bc} = e^{\nu}_{~b} \Bigl( e_{\nu c ,\mu} - \Gamma^{\rho}_{~\mu\nu}
e_{\rho c} \Bigr) \qquad , \qquad \Gamma^{\rho}_{~\mu\nu} = \frac12
g^{\rho\sigma} \Bigl(g_{\sigma \mu , \nu} + g_{\nu \sigma , \mu} -
g_{\mu\nu , \sigma} \Bigr) \; . \label{spincon}
\end{equation}
The gamma matrices $\gamma^a_{~ij}$ obey the usual anti-commutation 
relation, and their commutator gives the spin generator,
\begin{equation}
\{ \gamma^a , \gamma^b \} = -2 \eta^{ab} I \qquad , \qquad J^{bc} \equiv
\frac{i}{4} [ \gamma^a , \gamma^b ] \; . \label{gammaM}
\end{equation}
Finally, we note the relation,
\begin{equation}
\gamma^a J^{bc} = \frac{i}{2} \gamma^{[a} \gamma^b \gamma^{c ]} - 
\frac{i}{2} \eta^{ab} \gamma^c + \frac{i}{2} \eta^{ac} \gamma^b \; ,
\label{spinrelation}
\end{equation}
where square-bracketed indices are anti-symmetrized.

Expressing the Lagrangian (\ref{GR+QED}) in ADM form obviously requires
an explicit formula for the vierbein. It is useful to compare the ADM
form (\ref{ADMmetric}) of $g^{\mu\nu}$ with the $3+1$ expansion of 
expression (\ref{vierbein}),
\begin{equation}
g^{\mu\nu} = -e^{\mu}_{~0} e^{\nu}_{~0} + e^{\mu}_{~k} e^{\nu}_{~k} \; .
\end{equation}
It follows that we can take \cite{Deser:1976ay},
\begin{equation}
e^{\mu}_{~0} = \left( \begin{matrix} \frac1{N} \\ \frac{N^m}{N} \end{matrix}
\right) \qquad , \qquad e^{\mu}_{~k} = \left( \begin{matrix}
0 \\ \mathcal{E}^{m}_{~k} \end{matrix} \right) \; , \label{ADMvierbein}
\end{equation}
where $\mathcal{E}^{m}_{~k}$ is the inverse of the purely spatial vierbein (the 
dreibein) which obeys $\gamma^{ij} = \mathcal{E}^{i}_{~k} \mathcal{E}^{j}_{~k}$. If we use $i$ 
for the spatial component of the local Lorentz index $a$, the full vierbein 
and its inverse are,
\begin{equation}
e_{\mu a} = \left( \begin{matrix} -N & -\mathcal{E}_{ik} N^k \\ 0 & \mathcal{E}_{mi} \end{matrix} 
\right) \qquad , \qquad e^{\mu}_{~a} = \left( \begin{matrix} \frac1{N} & 0 \\
\frac{N^m}{N} & \mathcal{E}^{m}_{~i} \end{matrix} \right) \; . \label{bothv}
\end{equation}

There are three gauge symmetries to fix. For general coordinate invariance
we first write the 3-metric in terms of a graviton field $h_{ij}(t,\vec{x})$,
\begin{equation}
\gamma_{ij} = \Bigl[ e^{h} \Bigr]_{ij} = \delta_{ij} + h_{ij} + \frac12 
h_{ik} h_{kj} + \dots \; . \label{3metric}
\end{equation}
Then we impose the transverse-traceless conditions,
\begin{equation}
h_{ii}(t,\vec{x}) = 0 \qquad , \qquad \partial_{j} h_{ji}(t,\vec{x}) = 0 \; .
\label{GRgauge}
\end{equation}
Note that tracelessness implies $\sqrt{\gamma} = 1$. Three of the local 
Lorentz gauge conditions are implied by (\ref{ADMvierbein}). We use the 
remaining three to make the dreibein symmetric, so that the full set of 
local Lorentz gauge conditions is,
\begin{equation}
e_{m 0} = 0 \qquad , \qquad \mathcal{E}_{m i} = \mathcal{E}_{i m} \; , \label{localLgauge}
\end{equation}
Like Lorentz-symmetric gauge \cite{Woodard:1984sj}, this choice is also
ghost-free. Note that symmetry of the spatial part allows us to express
the dreibein in terms of the graviton field (whose indices are raised
and lowered with $\delta_{ij}$),
\begin{equation}
\mathcal{E}_{ij} = \Bigl[ e^{\frac12 h}\Bigr]_{ij} \qquad , \qquad 
\mathcal{E}^{ij} = \Bigl[ e^{-\frac12 h} \Bigr]_{ij} \; . \label{dreibein}
\end{equation}
Finally, there is the electromagnetic gauge. The most convenient choice
for us is an analogue of Coulomb gauge,
\begin{equation}
\partial_{i} \Bigl[ \frac{\gamma^{ij}}{N} \Bigl( \dot{A}_{j} - F_{jk}
N^k\Big) \Bigr] = 0 \; . \label{EMgauge}
\end{equation}

We can now express the Lagrangian in ADM form. The gravitational and
electromagnetic parts have the same simple dependence on the lapse
as the cosmological analogue (\ref{Nform}). Their contributions to the
kinetic and potential energies are,
\begin{eqnarray}
K_{\rm GR} = E_{ij} E_{k\ell} \Bigl( \gamma^{ik} \gamma^{j\ell} -
\gamma^{ij} \gamma^{k\ell}\Big) & , & P_{\rm GR} = -
\mathcal{R} \; , \qquad \label{ADMGR} \\
K_{\rm EM} = 8\pi G \Bigl( F_{0i} \!-\! F_{ik} N^k\Bigr) \Bigl(
F_{0j} \!-\! F_{j\ell} N^{\ell} \Bigr) \gamma^{ij} & , & 
P_{\rm EM} = 4\pi G F_{ij} F_{k\ell} \gamma^{ik} \gamma^{j\ell} .
\qquad \label{ADMEM}
\end{eqnarray}
The fermionic contributions are not so simple. Because the variation 
with respect to $N$ gives the Hamiltonian constraint, which involves 
no time derivatives of the fermion, parts of the gauge-fixed Lagrangian
must be independent of $N$ \cite{Deser:1976ay},
\begin{eqnarray}
\lefteqn{\mathcal{L}_0 = \overline{\psi} \gamma^0 \Bigl[ i \partial_0 
\!-\! e A_0 \!-\! \frac12 A_{0 k\ell} J^{k\ell} + N^m \Bigl(i \partial_m
\!-\! e A_m \!-\! \frac12 A_{m k\ell} J^{k \ell} \Bigr) \Bigr] \psi }
\nonumber \\
& & \hspace{1.5cm} + \frac12 \overline{\psi} \mathcal{E}^{m}_{~k} \mathcal{E}^{n}_{~i}
\Bigl[ \dot{\gamma}_{mn} \!+\! N^{k} \gamma_{m n , k} \!+\! \gamma_{mk}
\partial_{n} N^k \!+\! \gamma_{n k} \partial_{m} N^k\Bigr] \gamma^{k}
J^{0i} \psi \; . \label{special} \qquad
\end{eqnarray}
The Dirac contributions to the kinetic and potential terms are,
\begin{eqnarray}
K_{\rm D} &\!\!\! = \!\!\!& 8\pi G \, \overline{\psi} \mathcal{E}^{m}_{~i} N^{k}
\Bigl[ \partial_{k} (\gamma_{m\ell} N^{\ell}) - \partial_{m} 
(\gamma_{k\ell} N^{\ell}) \Bigr] \tfrac{i}{2} \gamma^{i} \psi \; ,
\label{ADMDK} \\
P_{\rm D} &\!\!\! = \!\!\!& -16 \pi G \, \overline{\psi} \mathcal{E}^{m}_{~k} 
\gamma^{k} \Big( i\partial_{m} - e A_{m} - \frac12 A_{m ij} J^{ij}\Bigr) 
\psi + 16 \pi G \, m_0 \overline{\psi} \psi \; . \qquad \label{ADMDP}
\end{eqnarray}

As with the cosmological analogue of Section 4, the lapse and shift are
constrained variables which must be expressed in terms of the physical
degrees of freedom, $h_{ij}$, $A^{T}_{i}$ (the transverse vector 
potential which obeys (\ref{EMgauge})), $\overline{\psi}$ and $\psi$.
The same is true of $A_{0}$, which is determined by the Gauss's Law
constraint,
\begin{eqnarray}
0 &\!\!\! = \!\!\!& \frac{\delta S}{\delta A_0} = \partial_{\nu} \Bigl[ 
\sqrt{-g} \, g^{\nu\rho} g^{0 \sigma} F_{\rho\sigma}\Bigr] - e \overline{\psi} 
e^{0}_{~a} \gamma^{a} \psi \sqrt{-g} \; , \label{variation} \qquad \\
&\!\!\! = \!\!\!& \partial_{j} \Bigl[ \frac{\gamma^{jk}}{N} \Bigl( F_{0k}
- F_{k\ell} N^{\ell}\Bigr) \Bigr] - e \overline{\psi} \gamma^{0} \psi =
-\partial_{j} \Bigl[ \frac{\gamma^{jk} \partial_{k} A_{0}}{N}\Bigr] - 
e \overline{\psi} \gamma^{0} \psi \; . \label{Gauss}
\end{eqnarray}
Like the Momentum Constraint (\ref{Momentum}) of the cosmological analogue, 
this equation cannot be solved exactly for $A_0$ owing to the factor of 
$1/N$, which involves $A_0$ (through $K_{\rm EM}$), once one solves the 
Hamiltonian Constraint,
\begin{equation}
\frac1{N} = \sqrt{\frac{P_{\rm GR} + P_{\rm EM} + P_{\rm D}}{-K_{\rm GR}
- K_{\rm EM} - K_{\rm D}}} \; . \label{badfactor}
\end{equation}
We might also wish to reconsider the electromagnetic gauge condition
(\ref{EMgauge}) in order to give a more transparent specification of the
physical electromagnetic degrees of freedom. Another possibility would 
be to impose the simple gauge condition $A_0 = 0$ and use the constraint
equation to solve for the longitudinal part of $A_{i}$.

Although our analysis of this system is not yet complete, it is clear
that one can solve the Hamiltonian Constraint for the lapse, the same
as for the cosmological analogue of Section 4. This will produce a 
reorganized sort of perturbation theory because it mingles together the
negative gravitational degrees of freedom with the positive energy degrees
of freedom which source them. It is not clear whether or not that is
enough to realize Stanley Deser's dream of quantum gravity regulating
ultraviolet divergences. Issues which require further study are:
\begin{itemize}
\item{Deciding between Coulomb Gauge and Temporal Gauge, as already
discussed.}
\item{Understanding the role of the ``special'' part of the Lagrangian
(\ref{special}).}
\item{Choosing how to extract ``background'' parts of the kinetic and
potential energies so as to keep quantum corrections perturbatively 
small.}
\end{itemize}
The last issue is potentially the most important. For the cosmological
analogue of Section 4 one has,
\begin{eqnarray}
K &\!\!\! = \!\!\!& -2 (3 H^2 \!+\! \dot{H}) + 4 H \Delta E_{ij} \gamma^{ij}
+ \Delta E_{ij} \Delta E_{k\ell} \Bigl(\gamma^{ik} \gamma^{j\ell} \!-\!
\gamma^{ij} \gamma^{k\ell}\Bigr) \; , \qquad \\
P &\!\!\! = \!\!\!& 2 (3 H^2 \!+\! \dot{H}) - \mathcal{R} \; , \qquad 
\end{eqnarray}
where we define,
\begin{equation}
\Delta E_{ij} \equiv E_{ij} + H \gamma_{ij} \; .
\end{equation}
One consequence is that the lapse is unity at zeroth order. For QED $+$
GR we want to implement a similar expansion, and we want to exploit this
freedom to keep quantum corrections small.

\section{Conclusions}

Stanley Deser was a great physicist and a good man who left the world a
better place. Section 1 reviews some of his most important contributions
to physics while section 2 presents personal reminiscences from one of 
his students. The remainder of the paper is devoted to one motivation
for Deser's early fascination with quantum gravity: the possibility that
it might regulate its own divergences and those of other theories 
\cite{Deser:1957zz}. This possibility arises because the gravitational 
interaction energy is negative, and sourced by the same sectors which
diverge.

Section 3 reviews the example ADM discovered of how classical (that is, 
non-quantum) general relativity cancels the famous linear divergence of
a point charged particle \cite{Arnowitt:1960zza,Arnowitt:1960zz}. The
final result (\ref{ADM3}) is not only finite but also independent of the
bare mass, as long as that is finite. This raises the fascinating 
prospect of not only solving the problem of quantum gravity but also
computing fundamental particle masses from first principles. It is 
impossible to overstate the revolution this would work on our perception
of quantum gravity. From a sterile issue of logical consistency, without
observable consequences at ordinary energies, and only perturbatively 
small effects even at the fantastic scales of primordial inflation, 
quantum gravity would be thrust to center stage. Every measurement of 
a fundamental particle mass would represent a sensitive check. One
might even hope that the mysterious 2nd and 3rd generations of the
Standard Model emerged as excited states of the 1st generation.

Of course there is a catch: one must make the calculation 
nonperturbatively in an interacting quantum field theory. This is 
evident from how its classical limit (\ref{ADM3}) depends upon 
$\alpha$ and $G$. There seems little hope of ever being able to perform 
an exact computation in an interacting $3+1$ dimensional quantum 
field theory. What is needed instead is a way of reorganizing 
conventional perturbation theory so that the negative energy constrained 
degrees of freedom have a chance to ``keep up'' with the positive 
energy, unconstrained degrees of freedom. The key to this seems to be 
solving the Hamiltonian Constraint. Section 4 describes how one 
accomplishes just that in the theory of primordial inflation. 
Fittingly, the solution (\ref{constrained}) is given using ADM 
variables. Although it is not clear if this form regulates the usual 
ultraviolet divergences of gravity with a scalar \cite{tHooft:1974toh},
the weak field expansion (\ref{constL}) of the gauge-fixed and 
constrained action does show an ADM-like erasure of the scalar
perturbation except for those parts protected by the nonzero first
slow roll parameter $\epsilon$.

Section 5 represents our initial attempt to implement the ADM mechanism
for Quantum Electrodynamics $+$ General Relativity. Although it is clear
that the Hamiltonian Constraint can be solved exactly, a number of issues
remain, the most important of which is how to extract ``background'' parts
of the kinetic and potential energies so as to keep quantum corrections
small. We look forward to further study of this system.

\vskip .5cm

\centerline{\bf Acknowledgements}

RPW is grateful for a lifetime of conversation and collaboration with 
N. C. Tsamis. This work was partially supported by NSF grant PHY-2207514 
and by the Institute for Fundamental Theory at the University of Florida.

\end{document}